\documentclass[10pt,letterpaper]{article}
\usepackage[top=0.85in,left=2.75in,footskip=0.75in]{geometry}

\usepackage{amsmath,amssymb}

\usepackage{changepage}

\usepackage{textcomp,marvosym}

\usepackage{cite}

\usepackage{nameref,hyperref}

\usepackage[right]{lineno}

\usepackage[nopatch=eqnum]{microtype}
\DisableLigatures[f]{encoding = *, family = * }

\usepackage[table]{xcolor}

\usepackage{array}

\newcolumntype{+}{!{\vrule width 2pt}}

\newlength\savedwidth

\raggedright
\usepackage[aboveskip=1pt,labelfont=bf,labelsep=period,justification=raggedright,singlelinecheck=off]{caption}

\makeatletter
\renewcommand{\@biblabel}[1]{\quad#1.}
\makeatother

\usepackage{lastpage,fancyhdr,graphicx}
\usepackage{epstopdf}
\fancyheadoffset[L]{2.25in}
\fancyfootoffset[L]{2.25in}
\usepackage[acronym,xindy]{glossaries}

\DeclareMathOperator*{\argmax}{arg\,max}

\newcommand{\ntime}{n_t} 
\newcommand{\nobs}{n_y} 
\newcommand{\nstate}{n_x} 
\newcommand{\ntheta}{n_\theta} 
\newcommand{\nknownIn}{n_u} 

\newcommand{\knownIn}{u} 
\newcommand{\thetaopt}{\hat{\theta}} 
\newcommand{\lh}[1][\theta]{\mathcal{L}_\mathcal{D}(#1)} 
\newcommand{\nllh}[1][\theta]{\mathcal{J}(#1)} 
\newcommand{\pl}[1][\ipar]{\text{PL}_\ipar(\xi)}
\newcommand{\labelsites}{m_{X_i}}

\newcommand{\itime}{k} 
\newcommand{\iobs}{j} 
\newcommand{\ipar}{k} 

\newcommand{\ymeas}{\bar{y}} 

\newcommand{\Rspace}[1]{\mathbb{R}^{#1}}

\newacronym{tag}{TAG}{Triacylglycerole}
\newacronym{dag}{DAG}{Diacylglycerole}
\newacronym{mag}{MAG}{Monoacylglycerole}

\begin{document}
\vspace*{0.2in}

\begin{flushleft}
{\Large
\textbf{Balancing label resolution and computational cost in dynamical models of lipid metabolism
}}
\newline
\\
Paul Jonas Jost\textsuperscript{1,2},
Christoph Thiele\textsuperscript{2},
Jan Hasenauer\textsuperscript{1,2*}
\\
\bigskip
\textbf{1} Bonn Center for Mathematical Life Sciences, University of Bonn, Bonn, Germany
\\
\textbf{2} Life and Medical Sciences (LIMES) Institute, University of Bonn, Bonn, Germany
\\
\bigskip

%
%





* jan.hasenauer@uni-bonn.de

\end{flushleft}


%
\section*{Abstract}

Lipid metabolism is a central biological process that is commonly studied using destructive mass-spectrometry experiments. A recently proposed strategy, uses multiple labels to extract temporal information about lipid metabolism from a single destructive measurement. However, the computational complexity of the model-based data analysis increases rapidly with the number of labels, creating a fundamental trade-off between the information content of the measurements and the cost of analysis. Here, we examine how the number of modelled labels affects parameter estimation accuracy, trajectory recovery, and computational cost, and whether modelling fewer labels than are experimentally available can mitigate this trade-off. Using synthetic data from a five-label experiment, we find that modelling three of the five labels provides a practical balance between experimental feasibility, inferential power, and computational tractability. In an application to hepatocyte triglyceride cycling, we further show that the most cost-efficient, single-label model can yield biologically implausible predictions for unobserved species, whereas models that resolve more labels better constrain these latent dynamics. These results provide practical guidance for selecting model resolution in multi-label experiments and establish a quantitative basis for balancing inferential power against computational cost.


%
\section*{Author summary}

Lipid metabolism is highly dynamic: lipid species are continuously synthesised, modified, stored, and degraded. Multi-label lipidomics data can help quantify these processes, but the corresponding mathematical models grow combinatorially with the number of labels: a lipid species with multiple fatty-acid chains admits a number of labelled forms that scales polynomially in the number of labels, which can quickly render the model intractable. We asked how many labels are needed for reliable inference, and whether coarse-grained models with fewer explicitly represented labels can be used instead. By comparing models with different label resolutions, we found that most of the gain in parameter accuracy was already achieved with a small number of explicitly modelled labels. For the lipid systems considered here, multi-label experiments with up to five labels, modelling three of them provided a useful compromise between accuracy and computational cost. However, modelling fewer labels also means a less detailed mechanistic description of the system, which can leave parts of the system poorly constrained: a model may agree well with the observed data while making misleading predictions for hidden, unmeasured parts. Our results suggest that lower label resolution can be a viable way to control computational cost, but that the choice should be checked against predictions for unmeasured species, not just fit to the observed data.


\section*{Introduction}


Dynamical modelling has become an indispensable tool in systems biology, enabling mechanistic insights into intra- and inter-cellular processes across diverse biological contexts~\cite{Kitano2002}. Applications range from the study of intracellular signal transduction and gene regulation to metabolism (see, e.g.,~\cite{DeJong2002, AldridgeBre2006, LakrisenkoWei2021}). By providing a quantitative description of biological processes, dynamic models allow researchers to analyse system behaviour and test mechanistic hypotheses.

A key strength of dynamical models is their ability to describe how biological systems evolve over time and respond to perturbations. This descriptive power comes from kinetic parameters that are typically unknown and must be estimated from data. Realising the potential of dynamical models therefore requires data that are informative about the underlying dynamics. Time-resolved measurements are particularly important, as they reveal dynamic features that help distinguish between otherwise similar parameter configurations, improving structural and practical identifiability~\cite{RaueBec2010, WielandHau2021structural}. Limited temporal information can therefore restrict the identifiability, interpretability, and predictive power of a model.

Many measurement techniques used in biology are inherently destructive, including sequencing for transcriptomics and mass spectrometry for metabolomics and lipidomics~\cite{HrdlickovaTol2017, IslamAry2017, ThomasFre2022, DaiShe2022, Shuken2023}. Because each measurement process consumes or substantially alters the sample, the same biological system often cannot be followed across multiple time points. Temporal dynamics must therefore be inferred from different samples, which can introduce experimental variability, batch effects, and other confounding factors that obscure the underlying dynamics~\cite{GohYon2022}. As a result, many studies remain limited to static analyses~\cite{KimPar2022}.

Several experimental strategies have been developed to make temporal information accessible despite these limitations. One strategy is to use non-destructive measurement techniques, such as fluorescence microscopy, nuclear magnetic resonance spectroscopy, and Raman spectroscopy, which allow the same system to be observed repeatedly over time~\cite{ShroffTes2024, KumarJai2026, WielandMas2021}. These techniques can provide direct temporal trajectories, but are often limited in molecular resolution compared with mass-spectrometry-based approaches~\cite{AretzMei2016}. A complementary strategy is therefore to retain the destructive readout but encode temporal information into the sample beforehand, for example by introducing distinguishable labels at defined time points. Labels, in this setting, are chemically distinguishable variants of a structural component of the molecule, e.g., for lipids, multiple specific fatty-acid chains. They are chosen to be metabolically equivalent so that they trace material without perturbing the reactions of interest. When a single molecule has multiple such sites, each can carry a label from a different time point, and the number of distinguishable labelled forms grows combinatorially with the number of labels. Such labelling strategies can generate quasi time series from a single measurement, which can then be decoded computationally~\cite{TrieblWen2018}. They thereby preserve access to rich molecular information, but require models capable of resolving multiple labelled species and their temporal relationships.
 
This creates a trade-off: more labels give more temporal information, but the number of labelled species to be tracked grows combinatorially with the number of incorporation sites. This is particularly acute for lipids with several fatty-acid chains, where explicitly label-resolved models can contain orders of magnitude more state variables than their unlabelled counterparts. Here, we focus on our recently introduced method for the study of lipid metabolism~\cite{JostWei2025}. The method introduces multiple metabolic labels at successive time points before a single destructive measurement, and reconstructs lipid dynamics from the resulting labelled species through model-based analysis. For this method an additional label can provide more information about the underlying dynamics, but also increases model size and therefore simulation and analysis cost. Indeed, the number and the temporal placement of labels may affect parameter identifiability, estimation accuracy, and computational feasibility. Without a quantitative understanding of these dependencies, experimental design and model construction remain largely heuristic, limiting the efficient use of such approaches in quantitative lipidomics. Rule-based modelling frameworks~\cite{harrisHog2016, BoutillierMaa2018} can automate the generation of these expanded reaction networks, but they do not alleviate the resulting simulation cost, which scales with the number of state variables. Thus, models that mirror the experimental labelling scheme one-to-one can become computationally prohibitive even for a moderate number of reactions, raising the question of whether the number of explicitly modelled labels must match the number of labels introduced experimentally, or whether modelling fewer labels can recover comparable inferential power at substantially lower computational cost.

In this study, we analyse the inference of lipid dynamics from multi-label lipidomics data using label-resolved dynamical models. We introduce a model-reduction strategy that maps data from experiments with multiple labels onto models with fewer explicitly represented labels, thereby partially decoupling experimental label number from model resolution. This enables a systematic comparison of computational cost, optimisation behaviour, parameter estimation accuracy, and trajectory recovery across label resolutions. Using synthetic data with known ground truth, we find diminishing returns in label resolution: most of the inferential gain is realized at moderate resolution, with further increases improving mainly yield more reliable convergence of the optimization, at additional computational cost. We then apply the framework to experimental hepatocyte triglyceride cycling data, demonstrating its practical use for analysing lipid dynamics from multi-label measurements.

\section*{Materials and Methods}

\subsection*{Mathematical modelling}

We consider a pulse-labelling experiment designed to extract temporal information from a destructive measurement. The general modelling formalism follows the label-based framework introduced in~\cite{JostWei2025}, adapted here to compare different levels of label resolution in the model. Cells are first maintained in unlabelled medium until they reach a metabolic steady state. Subsequently, distinguishable labels are introduced sequentially at prescribed time points. Labels are assumed to act as passive tracers: a labelled molecule participates in exactly the same reactions, with the same kinetics, as its unlabelled counterpart. Thus, labelled and unlabelled variants of biochemical species are treated as equivalent, except that they can be distinguished in the measurement.

Let $L$ denote the number of labels represented in the model. Labels are introduced at time points
\[
    T_1 < T_2 < \cdots < T_L,
\]
and we measure time relative to the first label introduction, i.e., $T_1=0$. Additionally, $T_0\leq T_1$ denotes the start of the experiment as a whole and $T_{L+1}\geq T_L$ denotes the time of measurement. In general, these time points can be chosen freely. However, in order to generate a pseudo-time trajectory, the labels need to be introduced at equidistant intervals~\cite{JostWei2025}. We denote the time between two consecutive label introductions by $t_{\text{pulse}}$. The time at which a label is introduced is an essential part of the model, because it bounds the window during which that label can be taken up from the medium.

\subsubsection*{Base model}

We first define a base model that describes the biochemical reaction network without resolving labels. The base model consists of biochemical species $X_i$, $i=1,\dots,n_X$, and reactions
\[
    R_k:\quad \sum_{i=1}^{n_X} S^-_{i,k}X_i \longrightarrow \sum_{i=1}^{n_X} S^+_{i,k}X_i,
    \qquad k=1,\dots,n_R,
\]
with stoichiometric coefficients $S^-_{i,k}$ and $S^+_{i,k}$. The reaction rates may depend on the species concentrations and on unknown model parameters.

The corresponding dynamical model is formulated as a system of ordinary differential equations (ODEs). Let $[X](t,\theta,\knownIn)\in\Rspace{n_X}$ denote the vector of concentrations of the biochemical species at time $t$, where $\theta\in\Rspace{\ntheta}$ denotes the vector of unknown model parameters, such as reaction rates and scaling parameters, and $\knownIn\in\Rspace{\nknownIn}$ denotes the vector of model inputs that correspond to the experimental conditions. The temporal evolution of the base model is given by
\begin{align}
    \frac{d[X]}{dt} = F([X](t,\theta, \knownIn),\theta, \knownIn),
\end{align}
where $F:\Rspace{n_X}\times\Rspace{\ntheta}\times\Rspace{\nknownIn}\to\Rspace{n_X}$ denotes the right-hand side of the ODE system.

Before the introduction of labels, cells are assumed to reside in a metabolic steady state in unlabelled medium, i.e. $\tfrac{d[X]}{dt}=0$. This steady state provides the initial condition for the label-resolved dynamics. 

\subsubsection*{Label model}

To describe the dynamics of labelled resolved species, we derive a label model from the base model. The label model resolves, for each biochemical species, the possible labelled variants that differ in their incorporated labels but are otherwise governed by the same biochemical reaction rules. We use the term biochemical species for the entities in the base model, labelled variants for the label-resolved variants of these species, and state variables for the ODE variables describing their concentrations.

For a model with $L$ unique labels, we define the label set as
\begin{align}
    \mathcal{L}_L = \{0,1,\dots,L\},
    \label{eq:label_space}
\end{align}
where $0$ denotes the unlabelled variant and $1,\dots,L$ denote the distinguishable labels. Each biochemical species $X_i$ has a maximum number of label incorporation sites, denoted by $\labelsites$. The order and position of incorporated labels are not distinguished. A labelled variant of species $X_i$ is therefore described by a multi-index
\begin{equation}
    \ell\in
    \left\{
    \ell=(\ell_k)_{k=0}^{L}\in\mathbb{N}_0^{L+1}
    \,\Bigg|\,
    \sum_{k=0}^{L}\ell_k=\labelsites
    \right\},
    \label{eq:labels}
\end{equation}
where $\ell_k$ denotes how often label $k$ is incorporated. The corresponding state variable in the label model is denoted by $x_{i,\ell}(t,\theta, \knownIn)$. A labelled variant is called purely labelled if all incorporation sites carry the same label, i.e., if $\ell_k=\labelsites$ for some $k\in\{1,\dots,L\}$.

The label model is constructed such that labels are transferred according to the biochemical reactions of the base model, while the reaction kinetics remain label-\allowbreak independent. The availability of labelled material is determined by the experimental pulse schedule: before label $k$ is introduced at time $T_k$, no new molecular forms containing label $k$ can be generated through uptake from the medium; after $T_k$, label $k$ is available as the externally supplied label until at $T_{k+1}$, it is replaced by label $k+1$. Thus, we define the model input vector as a scaling factor for the reaction influx of labels $\knownIn:\mathbb{R}\to\Rspace{L+1}$ component-wise by $u(t) = \bigl(u_0(t), u_1(t), \ldots, u_L(t)\bigr)^\top$, where
\begin{equation}
u_0(t) =
\begin{cases}
1, & t \notin [T_1, T_L), \\
0, & t \in [T_1, T_L),
\end{cases}
\end{equation}
and for $i = 1, \ldots, L$,
\begin{equation}
u_i(t) =
\begin{cases}
1, & t \in [T_i, T_{i+1}), \\
0, & \text{otherwise},
\end{cases}
\end{equation}
In this way, the pulse schedule links the time of label introduction to the generation of labelled variants.

To describe the dynamics of the label model, let $x(t,\theta)\in\Rspace{\nstate}$ denote the vector collecting all state variables $x_{i,\ell}(t,\theta)$. Its dynamics are described by
\begin{align}
    \frac{dx}{dt} = f(x(t,\theta),\theta), 
    \qquad
    x(0,\theta)=x_0(\theta),
    \label{eq:label_ode}
\end{align}
where $f:\Rspace{\nstate}\times\Rspace{\ntheta}\to\Rspace{\nstate}$ is obtained from the base reaction network by accounting for all admissible label transfers. The initial condition $x_0(\theta)$ represents the metabolic steady state before the first label is introduced. Accordingly, all label-resolved species containing non-zero labels are initially zero, while the unlabelled variants are initialised according to the steady state of the base model. Since summing the label-resolved in- and outfluxes of each metabolite over all label configurations recovers the fluxes of the base model, the model remains in metabolic steady-state throughout the whole experiment, i.e.
\begin{equation}
    \sum\limits_{\ell}\frac{dx_{i,\ell}}{dt} = 0.
\end{equation}
Thus, the total concentration of each biochemical species, summed over all labelled and unlabelled variants, remains equal to the corresponding steady-state concentration of the base model.

\subsubsection*{Label-dependent model growth}

The number of state variables in the label model increases with the number of explicitly represented labels. For a biochemical species $X_i$ with $\labelsites$ label incorporation sites, the number of possible labelled variants is the number of ways to distribute $\labelsites$ indistinguishable incorporation sites across $L+1$ label categories. This yields
\begin{equation}
    \binom{L+\labelsites}{\labelsites}
    =
    \prod_{k=1}^{\labelsites}\frac{L+k}{k}.
    \label{eq:polygrowth}
\end{equation}
The total number of state variables in the label model is obtained by summing Eq–\eqref{eq:polygrowth} over all biochemical species in the base model. Thus, for fixed incorporation capacities, the number of state variables grows polynomially in $L$, with degree determined by the maximum number of incorporation sites across species. This growth can be substantial: a single species with three labelling sites already requires 56 state variables to explicitly represent L=5 labels, and lipid species with several fatty-acid chains or base networks with many biochemical species compound the effect.

\subsubsection*{Observables and measurement model}

Observables are measurable quantities derived from the state variables of the label model. In this study, observables correspond to selected labelled variants or sums over labelled variants. We denote the observables by $y(t,\theta)\in\Rspace{\nobs}$, where $\nobs$ is the number of observables considered. For the cases considered here, the observation function is linear and can be written as
\begin{align}
    y(t,\theta) = Hx(t,\theta),
    \label{eq:obs}
\end{align}
where $H\in\{0,1\}^{\nobs\times\nstate}$ is a selection or aggregation matrix. Each row of $H$ defines one observable by selecting or summing the corresponding state variables. 

Measurement data are noise-corrupted assessments of the observables at known time points, $\mathcal{D} = \{(t_\itime, \{\ymeas_{\iobs,\itime}\})\}_{\iobs,\itime}^{\nobs,\ntime}$. For the purpose of this study, we considered multiplicative log-normal noise,
\begin{align}
    \ymeas_{\iobs,\itime}
    =
    y_\iobs(t_\itime,\theta)\varepsilon_{\iobs,\itime},
    \qquad
    \log \varepsilon_{\iobs,\itime}
    \sim
    \mathcal{N}(0,\sigma_{\iobs,\itime}^2),
    \label{eq:noise_model}
\end{align}
where $\sigma_{\iobs,\itime}$ denotes the standard deviation of the log-transformed measurement error.

\subsection*{Model reduction via label downshifting}

As the number of state variables in the label model grows rapidly with the number of explicitly represented labels, reduced models can be useful for assessing part of the information contained in a multi-label experiment at lower computational cost. In general, such reductions lead to discardment of measurements that cannot be represented in the reduced model. Here, we propose a family of reduced label models that avoids this direct loss of data by mapping measurements from an experiment with many labels onto a model with fewer explicitly represented labels.

The construction exploits the sequential pulse structure of the experiment. In principle, we treat the first label identical to no label: If labels are introduced at equidistant time points and behave identically in the biochemical network, then the dynamics associated with earlier and later labels differ primarily by the time at which the corresponding label became available. Measurements involving an earlier label can therefore be interpreted, under the assumptions of the label model, as time-shifted information about a reduced label system~\cite{JostWei2025}. The resulting label downshifting procedure removes the earliest label from the explicit model representation, merges it with the unlabelled index, shifts the remaining label indices, and adjusts the associated measurement times.

We first consider the reduction from a model with $L$ explicitly represented labels to a model with $L-1$ labels. The label space Eq–\eqref{eq:label_space} is mapped from $\mathbb{N}_0^{L+1}$ to $\mathbb{N}_0^{L}$. For a labelled variant with multi-index $\ell\in\mathbb{N}_0^{L+1}$, we define
\begin{align}
    \varphi_{L\to L-1}:\mathbb{N}_0^{L+1}\to\mathbb{N}_0^{L}.
\end{align}
For labelled variants that are not purely labelled with label $1$, the transformed multi-index $\ell'=\varphi_{L\to L-1}(\ell)$ is given by
\begin{align}
    \ell'_0 &= \ell_0+\ell_1,&\ell'_k &= \ell_{k+1},&&&k=1,\dots,L-1,&\qquad\text{if } \ell_1\neq \labelsites,\nonumber\\[2mm]
    \ell'_0 &= 0,&\ell'_1 &= \ell_1,&\ell'_k &= 0,&k=2,\dots,L-1,&\qquad\text{if } \ell_1= \labelsites.
    \label{eq:label_downshift}
\end{align}
Thus, label $1$ is merged with the unlabelled category, while labels $2,\dots,L$ are relabelled as $1,\dots,L-1$. Purely labelled variants containing only label $1$ are the only exception: their label identity is retained in the reduced representation because they correspond to time-shifted measurements of the other purely labelled variants (see Eq~\eqref{eq:downshift_rule}).

\begin{figure}[t!]
    \centering
    \includegraphics[width=\linewidth]{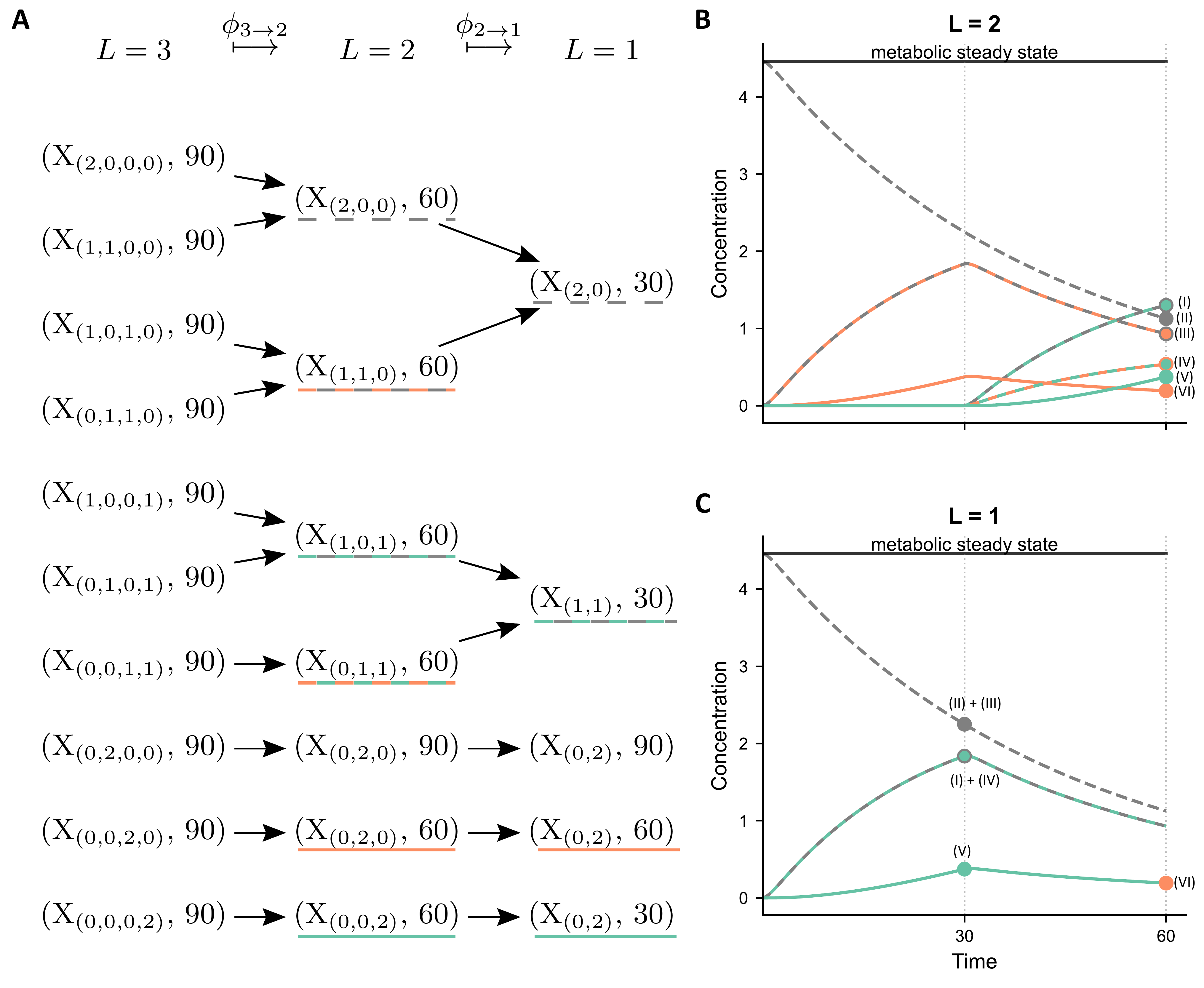}
    \caption{\textbf{Label downshifting.}
    (A)~The downshifting function $\phi$ illustrated on a species with $\labelsites=2$ label incorporation sites (see Eq–\eqref{eq:labels} for index notation) with measurement time point $t=90$, mapping all different label combinations from three labels to two and two to one, reducing the number of combinations from ten to seven to five.
    (B)~Dynamics of the labelled versions of a species with $\labelsites=2$ and $L=2$ distinct labels. Each colored line is a labelled species, matching the colors in (A); the black line is the sum over all species, confirming metabolic steady state. Roman numerals identify individual data points.
    (C)~Downshifted version of (B) to $L=1$: label 2 (orange) is relabelled as label 1, and species mapped to the same downshifted species and timepoint are summed. The purely labelled data point of the original label 1 (green) is retained as additional temporal information.}
    \label{fig:downshifting}
\end{figure}

The downshift rule acts jointly on label indices and measurement times. For a labelled variant of biochemical species $X_i$ with multi-index $\ell$ measured at time $t$, we define
\begin{align}
    \phi_{L\to L-1}:
    \mathbb{N}_0^{L+1}\times\mathbb{R}
    \to
    \mathbb{N}_0^{L}\times\mathbb{R}
\end{align}
by
\begin{align}
    \phi_{L\to L-1}(\ell,t)
    =
    \begin{cases}
        \big(\varphi_{L\to L-1}(\ell),\,t-t_{\text{pulse}}\big),
        & \text{if } \ell_1\neq \labelsites,\\[1mm]
        \big(\varphi_{L\to L-1}(\ell),\,t\big),
        & \text{if } \ell_1= \labelsites.
    \end{cases}
    \label{eq:downshift_rule}
\end{align}
Hence, measurements are shifted backward by one pulse interval unless they correspond to a purely labelled variant of the first label (Fig~\ref{fig:downshifting}A).

To reduce from $L$ labels to $L'<L$ labels, the one-step downshift transformation is applied repeatedly:
\begin{align}
    \phi_{L\to L'}
    =
    \phi_{L'+1\to L'}
    \circ
    \phi_{L'+2\to L'+1}
    \circ
    \cdots
    \circ
    \phi_{L\to L-1}.
    \label{eq:iterated_downshift}
\end{align}

Distinct observable--timepoint pairs in the original label model may map to the same observable--timepoint pair in the reduced model. In this case, their measurements are aggregated. Denoting by $\ymeas^{L'}_{\iobs',\itime'}$ the measurements used for the reduced $L'$-label model, we define
\begin{align}
    \ymeas^{L'}_{\iobs',\itime'}
    =
    \sum_{(\iobs,\itime)\in
    \phi_{L\to L'}^{-1}(\iobs',\itime')}
    \ymeas^{L}_{\iobs,\itime},
    \label{eq:measurement_downshift}
\end{align}
where $\phi_{L\to L'}^{-1}(\iobs',\itime')$ denotes the set of observable--timepoint pairs in the original label model that map to $(\iobs',\itime')$ under the downshift rule (Fig~\ref{fig:downshifting}B,C).

\subsection*{Parameter estimation}

For each considered label resolution, model parameters were inferred from the corresponding measurement dataset. Models with reduced label resolution generally use a transformed subset or aggregation of the measurements from the full multi-label experiment, as defined by the downshift rule in Eq–\eqref{eq:measurement_downshift}. Thus, each model is fitted only to the data that can be represented within its respective label space. Parameter estimation and uncertainty analysis were then performed independently for each label resolution.

Maximum likelihood (ML) estimates of the unknown model parameters were obtained by optimization. Based on the log-normal measurement model in Eq–\eqref{eq:noise_model}, and assuming independent measurement errors across observables and time points, the likelihood is
\begin{equation}
\lh
=
\prod_{\itime=1}^{\ntime}
\prod_{\iobs=1}^{\nobs}
\frac{1}{\sqrt{2\pi\sigma_{\iobs,\itime}^2}}
\frac{1}{\ymeas_{\iobs,\itime}}
\exp\left[
-\frac{1}{2}
\left(
\frac{
\log \ymeas_{\iobs,\itime}
-
\log y_\iobs(t_\itime,\theta)
}{
\sigma_{\iobs,\itime}
}
\right)^2
\right].
\label{eq:likelihood}
\end{equation}
The ML estimate is defined as
\begin{equation}
    \thetaopt
    =
    \argmax_{\theta\in\Rspace{\ntheta}}
    \lh.
    \label{eq:mle}
\end{equation}
For numerical optimisation, we equivalently minimise the negative log-likelihood,
\begin{equation}
    \nllh
    =
    -\log\lh
    =
    \frac{1}{2}
    \sum_{\itime=1}^{\ntime}
    \sum_{\iobs=1}^{\nobs}
    \left[
    \log\left(2\pi(\sigma_{\iobs,\itime} \ymeas_{\iobs,\itime})^2\right)
    +
    \left(
    \frac{
    \log \ymeas_{\iobs,\itime}
    -
    \log y_\iobs(t_\itime,\theta)
    }{
    \sigma_{\iobs,\itime}
    }
    \right)^2
    \right].
    \label{eq:nllh}
\end{equation}

The optimisation problem is generally non-convex and has multiple local optima. We therefore used multi-start optimisation with gradient-based local searches. In each start, parameters were initialised from different points in parameter space and subsequently refined by local optimisation of Eq–\eqref{eq:nllh}. The best parameter vector across all starts was used as the ML estimate.

Uncertainty in parameter estimates was assessed using profile likelihood analysis. For a fixed value $\theta_\ipar=\xi$ of parameter $\theta_\ipar$, the profile likelihood is defined as
\begin{equation}
    \pl
    :=
    \max_{\theta\,|\,\theta_\ipar=\xi}
    \lh
    =
    \exp\left(
    -
    \min_{\theta\,|\,\theta_\ipar=\xi}
    \nllh
    \right).
    \label{eq:profile_likelihood}
\end{equation}
In practice, $\theta_\ipar$ is varied over a prescribed range, and for each fixed value all remaining parameters are re-optimised. This yields a one-dimensional representation of the likelihood surface for each parameter while accounting for compensatory changes in the other parameters.

Under standard regularity conditions, the likelihood-ratio statistic associated with the profile likelihood follows a $\chi^2$ distribution with one degree of freedom~\cite{KreutzRau2013}. We define the profile likelihood--based confidence region for parameter $\theta_\ipar$ at confidence level $\alpha$ as
\begin{equation}
   \text{CR}^\alpha_\ipar
   =
   \left\{
       \xi\in\Rspace{}
       \,\Big|\,
       2\left(
       -\log\pl
       -
       \nllh[\thetaopt]
       \right)
       <
       \Delta_\alpha
   \right\},
   \label{eq:profile_confidence_region}
\end{equation}
where $\Delta_\alpha$ denotes the corresponding threshold of the $\chi^2_1$ distribution.

\subsection*{Implementation}

Model construction, simulation, parameter estimation, and uncertainty analysis were implemented using established tools for dynamic modelling and parameter inference. Parameter estimation problems were encoded in PEtab format~\cite{SchmiesterSch2021}. Simulations and sensitivity computations were performed with AMICI v0.34.2~\cite{FroehlichWei2021}. Parameter estimation was carried out with pyPESTO v0.5.8~\cite{SchaelteFro2023}. For the synthetic data analysis, optimisation used Fides v0.7.8~\cite{FroehlichSor2022fides} with $5000$ local starts per optimisation. For the experimental data analysis, optimisation used pyPESTO's implementation of the saCeSS algorithm~\cite{PenasGon2017} with a walltime limit of 47 hours and 8 workers.


\section*{Results}

To assess how label resolution affects model complexity, optimisation performance, parameter estimation, and dynamical predictions, we considered two complementary case studies. First, we used a synthetic benchmark derived from a triglyceride synthesis and cycling model~\cite{JostWei2025}. Because the ground-truth parameters and dynamics were known, this benchmark enabled a direct evaluation of parameter recovery and dynamic reconstruction. Second, we analysed an experimental hepatocyte triglyceride cycling dataset with three sequential labels, previously published in~\cite{JostWei2025}. This dataset allowed us to assess the practical impact of label resolution on the analysis of experimental multi-label lipidomics data; a model-based label-resolution analysis of these data had not been presented in the original study.

In both case studies, we compared models that differed in the number of labels represented explicitly. Reduced label models were obtained by mapping the available measurements to lower-dimensional label spaces, allowing us to study how much information is retained when model complexity is decreased.

\subsection*{Label downshifting decouples model resolution from experimental label number}

\begin{figure}[t!]
    \centering
    \includegraphics[width=\linewidth]{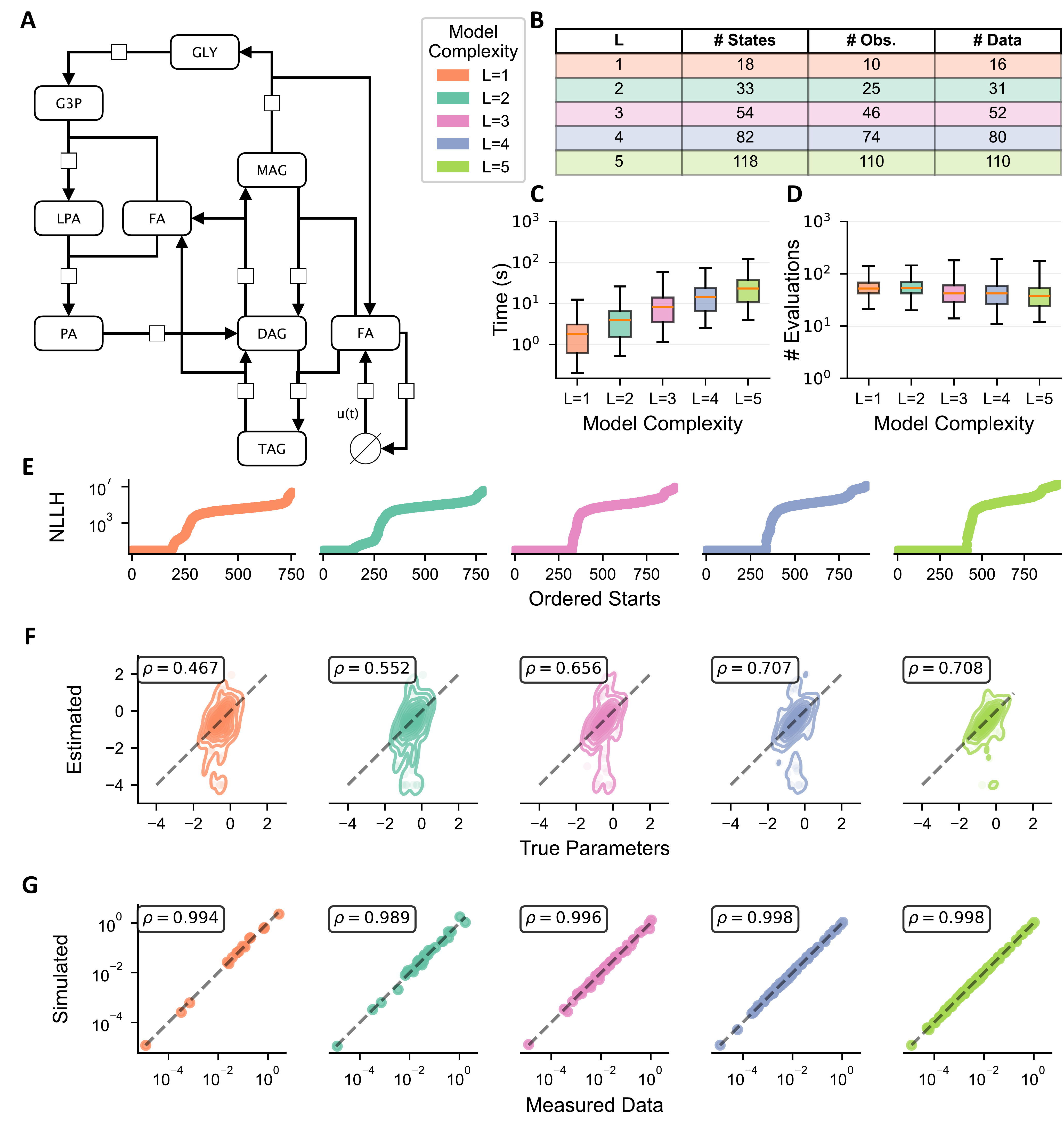}
    \caption{\textbf{Synthetic benchmark: number of state variables, and simulation / optimisation performance.}
    (A)~SBGN graph of the triglyceride synthesis and cycling benchmark model.
    (B)~Number of state variables, observables, and data points for each label model. Note that the base model corresponds to $L=0$.
    (C)~Simulation time for label models with different numbers of explicitly represented labels, evaluated for 10000 randomly sampled parameter vectors.
    (D)~Average number of objective-function evaluations per local optimisation start.
    (E)~Waterfall plot for one representative synthetic model, showing finite final objective-function values of individual optimisation starts.
    (F)~Parameter-estimation accuracy across 20 synthetic instances and identifiable parameter dimensions. $\rho$ denotes the spearman-rank correlation.
    (G)~Comparison of simulated and noisy measured observables across all synthetic instances.}
    \label{fig:toy_1}
\end{figure}

The complexity of label-resolved computational models increases rapidly with the number of labels, particularly when biochemical species can incorporate multiple labelled molecular components. We therefore first asked whether the dimensionality of the label model can be reduced without discarding the experimental setting itself. This would allow data from a multi-label experiment to be analysed with models of different complexity and thereby enable a direct assessment of the trade-off between computational cost and retained information.

We investigated this question using the synthetic triglyceride synthesis and cycling benchmark. The base model contained nine biochemical species, thirteen reactions, and thirteen parameters, including \acrfull{tag}, \acrfull{dag}, and \acrfull{mag} (Fig~\ref{fig:toy_1}A). Starting from a setup with $N_{\text{exp}}=5$ labels, we constructed reduced label models with $L=1,\dots,5$ explicitly represented labels. For each reduced label model, the data were mapped to the corresponding lower-dimensional label space using the downshifting approach described in the \textit{Materials and Methods} section, which builds on the time-shift property established in~\cite{JostWei2025}.

This procedure yielded a hierarchy of models based on the same biochemical reaction network but differing in the number of state variables and measurement points (Fig~\ref{fig:toy_1}B). We observed a reduction from 118 state variables to 18, accompanied by a reduction in the number of usable data points from 110 to 16. Importantly, reduced data points can represent aggregated measurements and may therefore combine information from several original readouts, potentially lowering measurement noise.

In summary, label downshifting enables a partial decoupling of experimental label number and model resolution. Lower-resolution models reduce the number of state variables and computational cost, but they use a coarser representation of the measured label information.

\subsection*{Simulation cost increases strongly with label resolution}

Having established the hierarchy of label models, we next assessed how label resolution affects computational cost for the synthetic triglyceride synthesis and cycling benchmark. For each model resolution, we measured simulation times using 2000 randomly sampled parameter vectors. We further created datasets from 20 randomly sampled parameter vectors (\nameref{S1_Table}) and optimised each (reduced) label model was fitted independently using multi-start optimisation with $5000$ local starts. We then analysed the number of objective-function evaluations required during local optimisation.

Mean simulation time increased more than tenfold between one (1.96s) and five (24.95s) explicitly represented labels (Fig~\ref{fig:toy_1}C). This increase followed the growth in state variables described above and reflects the higher cost of solving larger ODE systems. In contrast, the average number of objective-function evaluations per local optimisation start changed only moderately across label resolutions (Fig~\ref{fig:toy_1}D). Thus, the increase in total optimisation time was mainly driven by the higher simulation cost per evaluation, rather than by a substantially larger number of optimisation iterations.

The optimisation results further indicated that higher label resolution improved optimisation reproducibility. In the respective waterfall plots, models with more explicitly represented labels showed a more favourable distribution of final objective-function values across local starts (Fig~\ref{fig:toy_1}E). Thus, additional label-resolved information can improve optimisation behaviour in this benchmark, but this benefit comes at the cost of substantially more expensive simulations.

\subsection*{Parameter recovery improves but saturates with increasing label resolution}

We next asked whether the additional computational cost of higher label resolution translates into improved parameter estimation. The synthetic benchmark allowed this question to be assessed directly, because the ground-truth parameter values were known. Parameters previously identified as practically unidentifiable in~\cite{JostWei2025} were excluded from the parameter-recovery comparison to avoid confounding the analysis.

Parameter recovery improved as more labels were represented explicitly in the model (Fig~\ref{fig:toy_1}F). The improvement was most pronounced when moving from one to three labels (Spearman correlation increasing from 0.47 to 0.66). In contrast, the gain obtained by adding further labels became smaller at higher label numbers. Despite the differences in parameter accuracy, the fit to the measured observables was essentially indistinguishable across label resolutions, with Spearman correlation ranging between 0.994–0.998 (Fig~\ref{fig:toy_1}G).

These results indicate a saturation of inferential benefit for this synthetic benchmark. While representing more labels explicitly generally improved parameter recovery, most of the improvement was already achieved at intermediate label resolution. This suggests that, for this model and dataset, reduced label models can retain a large fraction of the information relevant for estimating identifiable parameters.

\begin{figure}[t!]
    \centering
    \includegraphics[width=\linewidth]{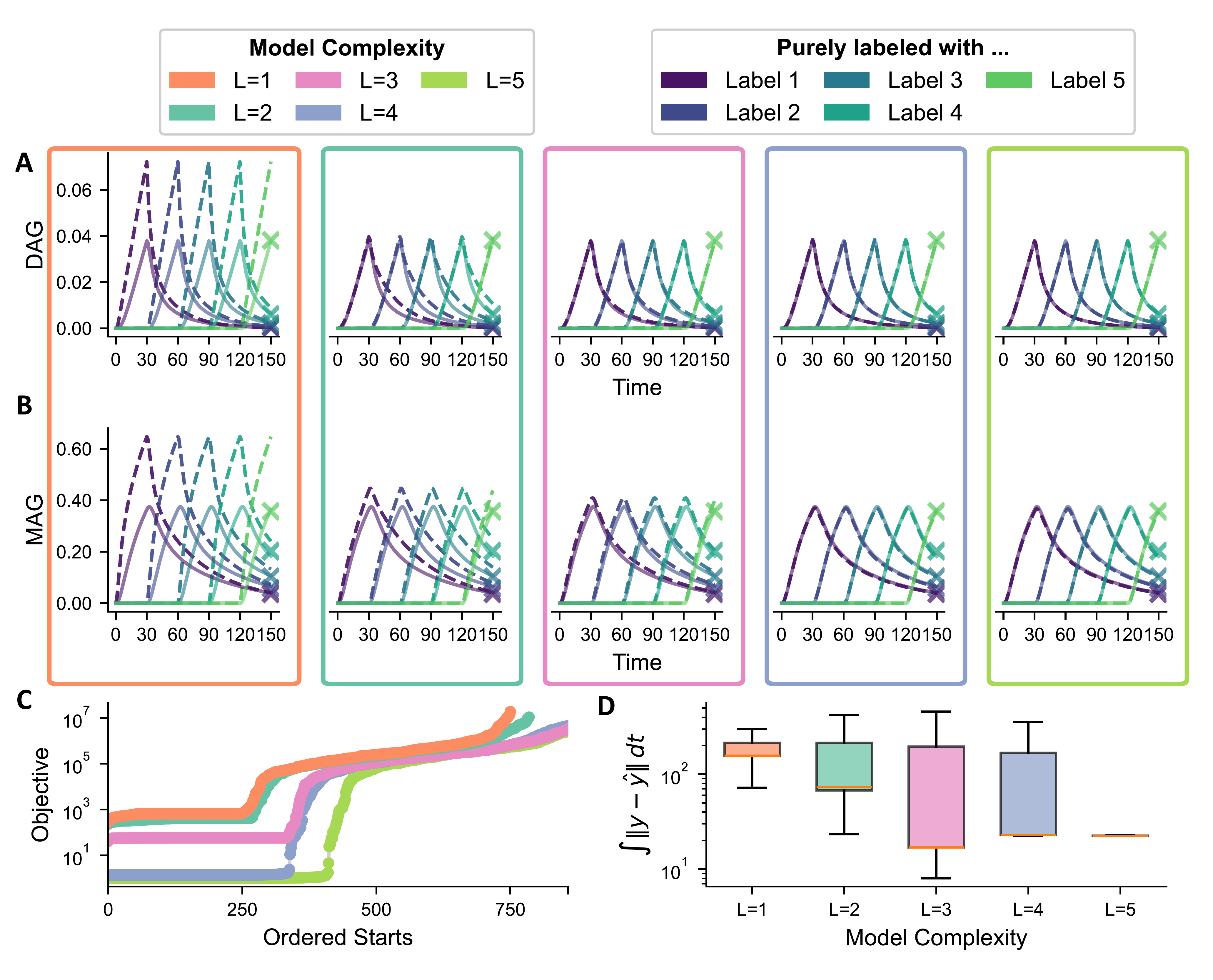}
    \caption{\textbf{Synthetic benchmark: reconstruction of underlying dynamics.}
    (A)~Dynamics of differently labelled \acrshort{dag} species. Solid lines show the ground-truth dynamics, dashed lines show simulations obtained with the best-fit parameters, and crosses indicate measurements.
    (B)~Dynamics of differently labelled \acrshort{mag} species, shown analogously to (A).
    (C)~Waterfall plot obtained by evaluating the objective function of the five-label model at all fitted parameter vectors from the different label resolutions.
    (D)~Accumulated deviation between simulated dynamics $\hat{y}$ and ground-truth dynamics $y$ for the best 500 parameter vectors from each label resolution.}
    \label{fig:toy_2}
\end{figure}

\subsection*{Intermediate label resolution recovers the underlying dynamics well in the synthetic benchmark}

Even for the full five-label model, some estimated parameters deviated substantially from the ground truth, reflecting practical non-identifiabilities in the benchmark problem (Fig~\ref{fig:toy_1}F). We therefore assessed not only parameter recovery, but also whether the fitted models reconstruct the underlying label-resolved dynamics. To this end, the best-fit parameters from each label resolution were evaluated in the five-label model and compared with the ground-truth trajectories used to generate the data.

As representative examples of the reconstructed label-resolved dynamics, we inspected the trajectories of \acrshort{dag} and \acrshort{mag}. For both species, agreement with the ground-truth dynamics improved as more labels were represented explicitly in the model (Fig~\ref{fig:toy_2}A,B). The one-label model showed pronounced deviations and tended to overestimate several labelled trajectories (up to 1.5 times the true concentration). In contrast, the differences between intermediate and high label resolutions were comparatively small for these examples. This behaviour was consistent with the global accumulated error across all species and over time: the one- and two-label models showed substantially larger errors, whereas intermediate and high label resolutions performed similarly well (Fig~\ref{fig:toy_2}D).

The comparison of fitted parameter vectors on the five-label objective function further showed that models with more labels tended to achieve lower objective values in the full label representation (Fig~\ref{fig:toy_2}C). At the same time, the accumulated-error analysis indicated that the best intermediate-resolution fits could reconstruct the true dynamics nearly as well as, and in some instances better than, the highest-resolution fits (Fig~\ref{fig:toy_2}D). This suggests that, in this benchmark, additional labels improve robustness and consistency, but the improvement in recovered dynamics saturates quickly. The results therefore point to a trade-off between computational cost, optimisation stability, and dynamical accuracy rather than to a universally optimal label number.

\subsection*{Experimental triglyceride cycling data are described by label-resolved models}

We next applied the modelling framework to experimental hepatocyte triglyceride cycling data with three sequential labels introduced at 30-minute intervals~\cite{JostWei2025}. This analysis served two purposes. First, it allowed us to assess whether the observations from the synthetic benchmark also translate to experimental multi-label lipidomics data. Second, it provided an opportunity to study the role of \acrshort{mag} as a potential intermediate in \acrshort{tag} synthesis and cycling. Although \acrshort{mag} is biochemically plausible as an intermediate, its concentration may remain low because of rapid turnover, making its involvement difficult to assess experimentally~\cite{WunderlingZur2023}.

\begin{figure}[t!]
    \centering
    \includegraphics[width=\linewidth]{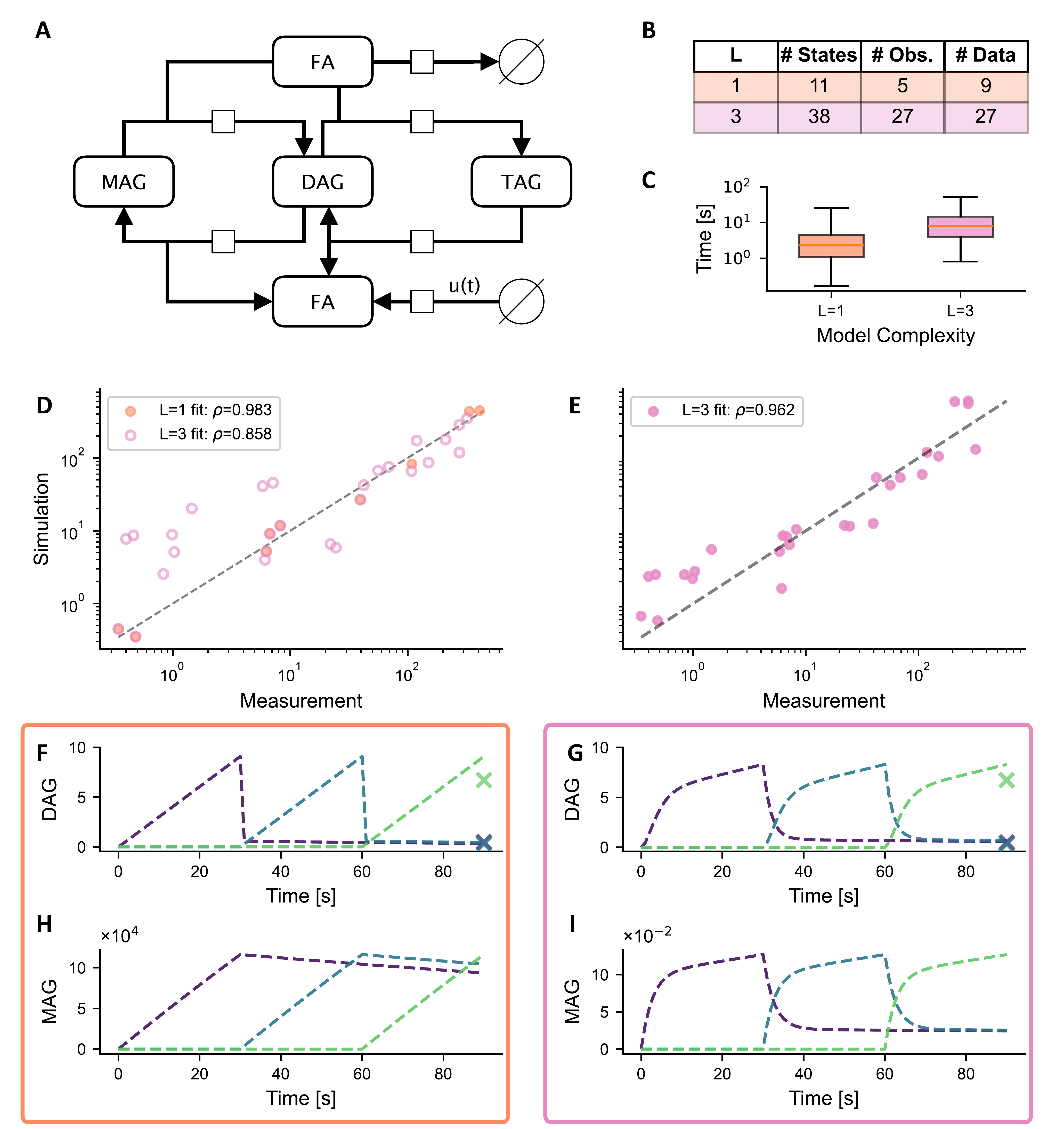}
    \caption{\textbf{Experimental triglyceride cycling model: fit and predicted dynamics.}
    (A)~SBGN graph of the triglyceride cycling model.
    (B)~Number of states, observables, and data points for each label model.
    (C)~Simulation time for the $L=1$ and $L=3$ model configurations, evaluated for 10000 randomly sampled parameter vectors.
    (D,E)~Comparison of measured and simulated observables for the $L=1$ and $L=3$ models. $\rho$ denotes the spearman-rank correlation. Fully colored dots represent the fit to data it was trained on. Outlined circles represent comparison of the parameters of the $L=1$ model against the $L=3$ data.
    (F,G)~Simulated dynamics of differently labelled \acrshort{dag} species for the $L=1$ and $L=3$ models. Dashed lines show simulations obtained with the best-fit parameters, and crosses indicate measurements.
    (H,I)~Predicted dynamics of differently labelled \acrshort{mag} species for the $L=1$ and $L=3$ models. \acrshort{mag} was not directly observed in the fitted dataset.}
    \label{fig:tg_overview}
\end{figure}

We constructed a label-resolved model of the proposed triglyceride cycling network and fitted it to the experimental data (Fig~\ref{fig:tg_overview}A, \nameref{S2_Table}). To assess the impact of label resolution, we compared a coarse one-label model ($L=1$) with the full three-label model ($L=3$). This comparison showed the expected computational trade-off: the average simulation time of the three-label model was approximately three times higher than that of the one-label model, consistent with its larger number of state variables (Fig~\ref{fig:tg_overview}B,C).

We then asked whether this additional computational cost improved the agreement with measured quantities and, more importantly, whether it affected predictions for unobserved intermediates. Our assessment of the model fits showed that both models achieved reasonable agreement with the respective datasets (Fig~\ref{fig:tg_overview}D,E). The one-label model provided a slightly worse fit to the $L=3$ data, but  still captured important aspects of the observed data.

The fitted models differed more strongly in their predicted latent dynamics than in their agreement with the measured data. For the observed \acrshort{dag} species, both models produced broadly comparable trajectories, although the three-label model followed the measured label-resolved dynamics more closely (Fig~\ref{fig:tg_overview}F,G). For the unobserved \acrshort{mag} species, the predictions diverged substantially (Fig~\ref{fig:tg_overview}H,I). The three-label model predicted low \acrshort{mag} concentrations, consistent with a rapidly turning-over intermediate, whereas the one-label model allowed \acrshort{mag} concentrations to increase to values far above the observed concentration scale of the species.

Thus, the experimental application shows that agreement with measured observables alone is insufficient to assess whether a reduced label model provides a reliable mechanistic description. In this case, the coarse model reproduced parts of the measured data but failed to constrain an unobserved intermediate to a biologically plausible range.

\begin{figure}[t!]
    \centering
    \includegraphics[width=\linewidth]{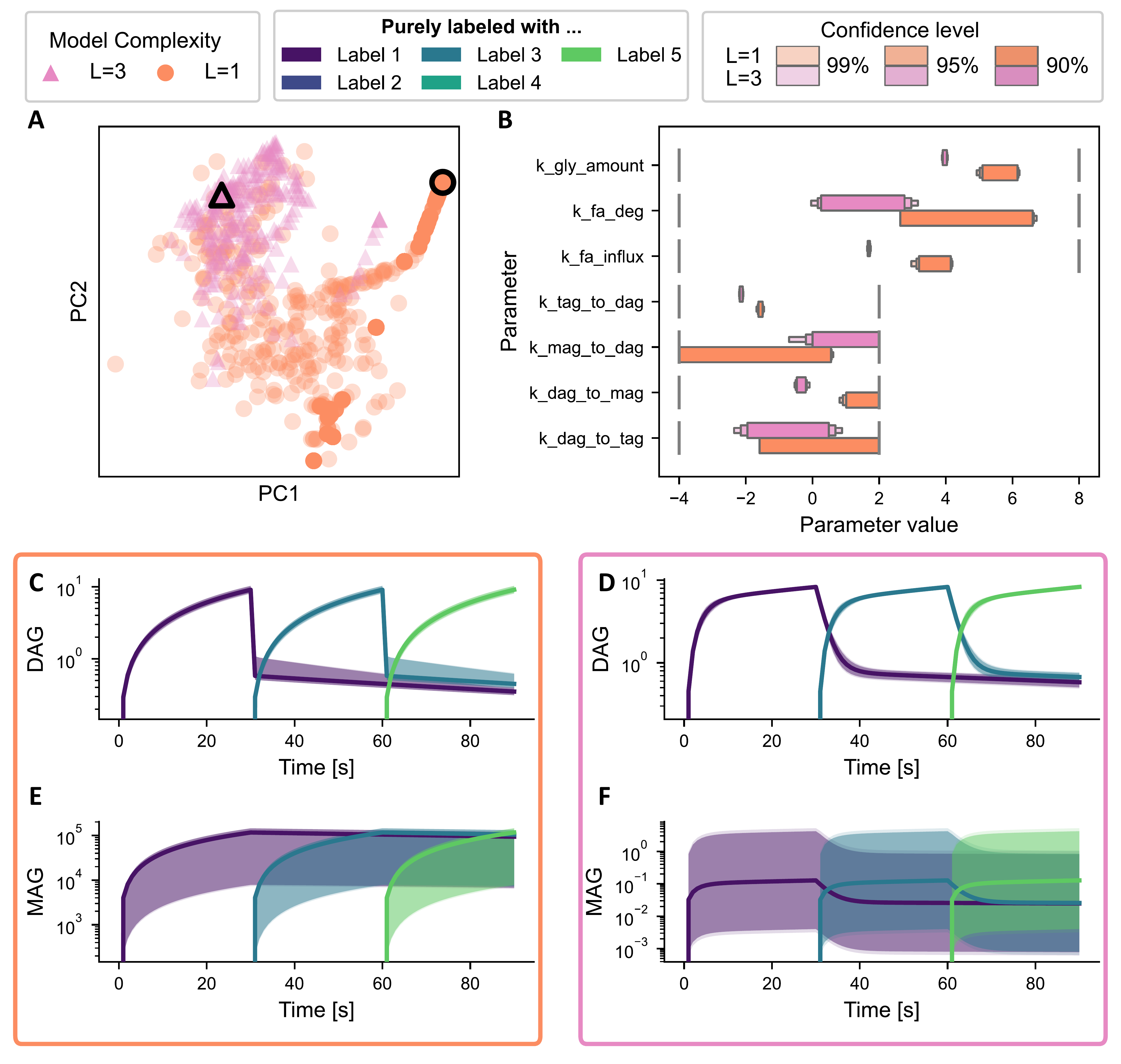}
    \caption{\textbf{Experimental triglyceride cycling model: optimisation results and uncertainty.}
    (A)~Principal component analysis of parameter vectors obtained from optimisation starts for the $L=1$ and $L=3$ models.
    (B)~Profile likelihood--based confidence intervals at 90\%, 95\%, and 99\% confidence levels for parameters of the $L=1$ and $L=3$ models.
    (C,D)~Uncertainty bands for \acrshort{dag} trajectories derived from profile-likelihood parameter uncertainties propagated to trajectories for the $L=1$ and $L=3$ models.
    (E,F)~Uncertainty bands for predicted \acrshort{mag} trajectories derived from profile-likelihood parameter uncertainties propagated to trajectories for the $L=1$ and $L=3$ models.}
    \label{fig:tg_unc1}
\end{figure}

\subsection*{Higher label resolution improves constraints on unobserved intermediates}

To investigate whether the divergent \acrshort{mag} predictions were caused by incomplete optimisation or by differences in parameter identifiability, we analysed the optimisation results and parameter uncertainties for the one- and three-label models (\nameref{S3_Figure}). In particular, we asked whether the one-label optimisation explored the parameter region associated with the three-label fit and whether the reduced model provided sufficient constraints on parameters controlling unobserved intermediates.

A principal component analysis of parameter vectors from the optimisation starts shows that the one-label optimisation explored parameter regions overlapping with those of the three-label model (Fig~\ref{fig:tg_unc1}A). The discrepancy between the fitted dynamics is therefore unlikely to be explained solely by a failure to initialise the one-label optimisation near the three-label solution. Instead, the reduced label resolution changed the information available for parameter estimation. Profile likelihood analysis showed broader confidence intervals for the one-label model for most parameters (Fig~\ref{fig:tg_unc1}B). In particular, differences in confidence intervalls of parameters associated with \acrshort{mag} were particulary large.

The trajectory uncertainty analysis showed that both models constrained the observed \acrshort{dag} dynamics comparatively well (Fig~\ref{fig:tg_unc1}C,D). In contrast, uncertainty for the unobserved \acrshort{mag} species spanned several orders of magnitude, especially in regions of parameter space that were weakly constrained by the data (Fig~\ref{fig:tg_unc1}E,F). Nevertheless, the three-label model kept the predicted \acrshort{mag} concentrations within a substantially lower and more biologically plausible range, whereas the one-label model allowed much larger values.

These findings indicate that the additional label-resolved information in the three-label model improves constraints on hidden parts of the reaction network. For this application, reducing the label resolution preserved much of the fit to observed species but weakened the ability to constrain unobserved intermediates.

\begin{figure}[t!]
    \centering
    \includegraphics[width=\linewidth]{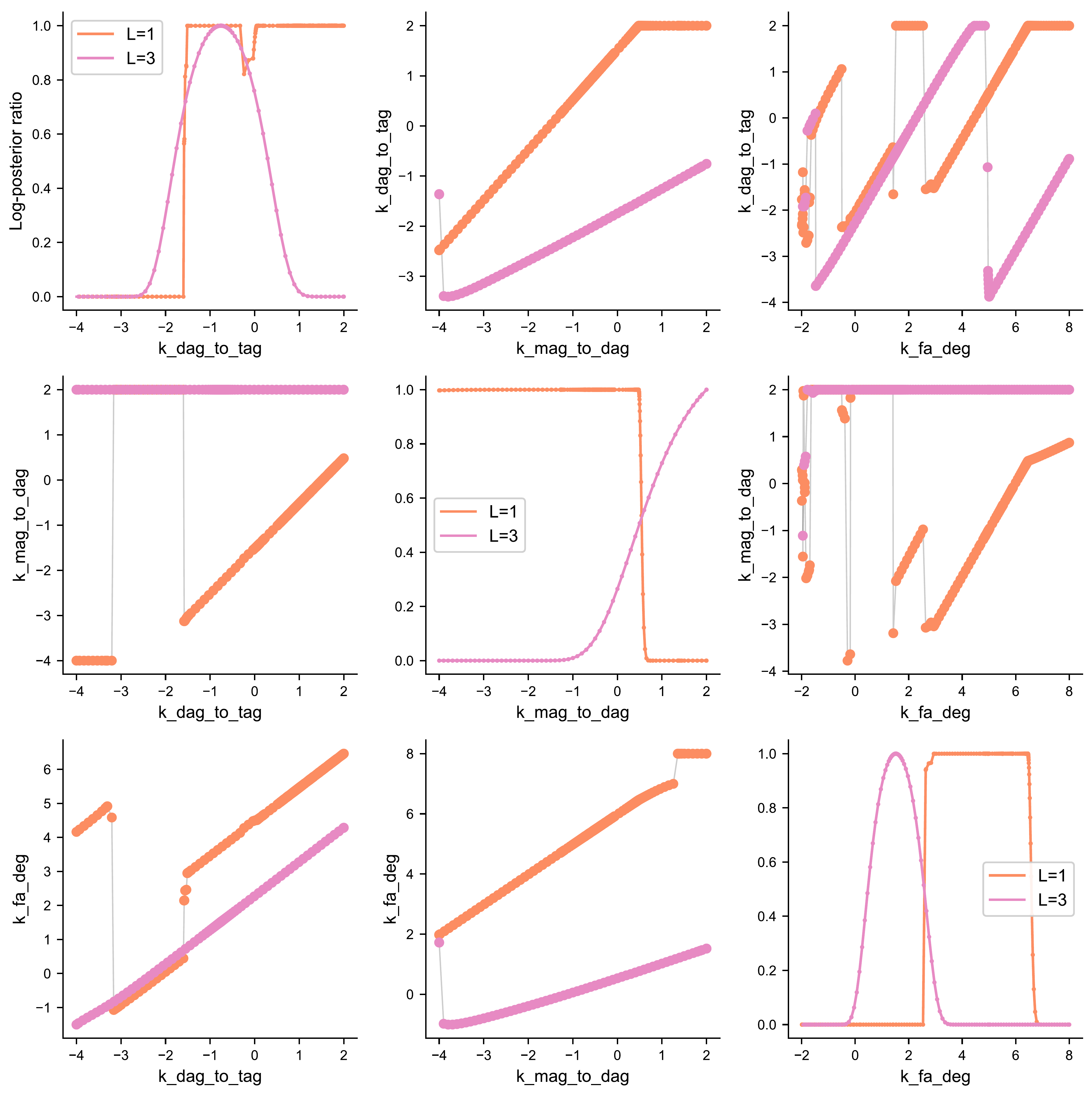}
    \caption{\textbf{Profile likelihood analysis for selected parameters in the experimental triglyceride cycling model.}
    Diagonal panels show profile likelihoods for selected parameters in the $L=1$ and $L=3$ models.
    Off-diagonal panels show re-optimised values of the row parameter as a function of the profiled column parameter, revealing compensatory parameter relationships and transitions between local optima.}
    \label{fig:tg_unc2}
\end{figure}

\subsection*{Profile likelihoods reveal stronger parameter dependencies in the reduced model}

Finally, we examined selected profile likelihoods in more detail to identify parameter dependencies that may explain the weaker constraints in the reduced model. After a complete overview (Figure in S3 Figure), we focused on parameters involved in \acrshort{dag}, \acrshort{mag}, and fatty-acid turnover, where the uncertainty analysis indicated substantial differences between label resolutions.

The profile likelihoods revealed pronounced compensatory relationships between several parameters (Fig~\ref{fig:tg_unc2}). These dependencies were more apparent in the one-label model, where changes in one profiled parameter could be partly compensated by re-optimising other parameters. Several profiles also showed transitions between local optima, visible as discontinuities or abrupt changes in the re-optimised parameter trajectories. Such behaviour is consistent with practical identifiability limitations in the reduced model.

For the three-label model, the profiles generally provided stronger constraints, although some confidence intervals still reached parameter bounds. Thus, the additional label resolution did not eliminate all identifiability challenges, but it reduced important ambiguities present in the one-label model. Together with the trajectory analysis, this supports the conclusion that higher label resolution can be important when unobserved intermediates are central to the mechanistic interpretation.

Overall, the synthetic and experimental analyses show that reduced label models can lower computational cost while retaining substantial information about measured quantities. In the synthetic benchmark, improvements saturated rapidly with increasing label resolution. In the experimental triglyceride cycling application, however, reduced label resolution led to weaker parameter constraints and stronger compensatory parameter dependencies, which mainly affected predictions for unobserved intermediates. The profile likelihood analysis further showed that the one-label model has stronger parameter dependencies and weaker practical identifiability than the three-label model, which constrained compensatory parameter changes more effectively and thereby reduced ambiguity in the inferred dynamics.


\section*{Discussion}

Multi-label experiments provide a way to recover temporal information from destructive measurements, but their model-based analysis can become computationally demanding. Our results demonstrate that the number of labels represented explicitly in the model can be reduced while retaining substantial information about the underlying dynamics. At the same time, the analysis shows that the appropriate model resolution depends on the intended use of the model, in particular on whether unobserved intermediates need to be constrained.

We introduced a downshifting approach that maps data from experiments with multiple labels onto models with fewer explicitly represented labels. This partially decouples the experimental number of labels from the model resolution used for analysis. This decoupling is practically relevant: an experiment with multiple labels can be conducted before the computationally optimal label resolution for analysis is determined, and the label resolution can be adjusted post-hoc without requiring new experimental data. In the synthetic five-label benchmark, reducing the number of modelled labels substantially lowered simulation cost, while a three-label resolution recovered most of the parameter and dynamic information obtained with the full label model. The improvement in parameter recovery and dynamic reconstruction saturated rapidly with increasing label number, suggesting that, for this benchmark, the main inferential benefit was achieved before all experimental labels were represented explicitly.

The application to experimental hepatocyte triglyceride cycling data showed that reduced label models can reproduce measured quantities while differing substantially in latent predictions. Both the one-label and three-label models achieved reasonable agreement with the observed data, but they led to different predictions for \acrshort{mag}, an unobserved intermediate in the considered reaction network. The three-label model constrained \acrshort{mag} concentrations to a range consistent with a rapidly turning-over intermediate, whereas the one-label model allowed substantially larger concentrations. This finding shows that fit quality for measured species alone is not sufficient to assess whether a reduced label model provides an adequate mechanistic description.

These results have direct implications for model construction in multi-label lipidomics. If the goal is to reproduce measured quantities or to obtain an initial assessment of parameter values, reduced label models may provide a useful and computationally efficient starting point. If the biological question depends on unobserved intermediates or detailed flux redistribution between labelled variants, higher label resolution may be required. Thus, the choice of model resolution should be guided not only by computational cost, but also by the mechanistic quantities that the model is expected to infer. For smaller models with fewer than one hundred state variables, we suggest modelling all labels, as computational cost is unlikely to be the bottleneck. For models with more than one hundred states, we suggest starting with a small test optimisation of one to three labels and assessing fit accuracy, since in the system we studied the accuracy gain after three labels decreases substantially.

Our study has several limitations. The synthetic benchmark was based on a relatively small reaction network and a fixed experimental design, and the observed saturation of inferential gain may not hold for larger networks, different pulse schedules, or models with stronger practical non-identifiabilities. The experimental analysis focused on one triglyceride cycling dataset with three labels, and the conclusions about \acrshort{mag} dynamics hinged on available prior knowledge. Broader benchmark collections for multi-label dynamical models would be needed to quantify accuracy--cost trade-offs across model classes and experimental designs, analogous to benchmark resources used in other areas of computational systems biology~\cite{HassLoo2019, Kreutz2019, VillaverdeFro2018}.

Future work should address both experimental design and computational implementation. Systematic studies across different lipid networks, pulse schedules, and measurement noise levels could provide more precise guidance for choosing the number and timing of labels in future experiments~\cite{NohNie2018, KangLi2025}. In parallel, automated pipelines for generating model hierarchies of different label resolution, together with simulation methods that exploit the repetitive pulse structure and combinatorial organisation of labelled species, could make larger label-resolved models tractable. Taken together, our findings support label-resolved dynamical modelling as a framework for extracting temporal information from destructive multi-label measurements and highlight the need to consider experimental labelling design, model granularity, and numerical efficiency jointly.

\section*{Supporting information}


\paragraph*{S1 Table}
\label{S1_Table}
\textbf{Parameter information of benchmark model.} A list of all parameters estimated in the benchmark model. Lower and upper bound denote the bound in the estimation process. Mean and standard deviation describe the distribution of the sampling processs for these parameters. Values are in $\log_{10}$, as the distribution followed a log-normal distribution.

\paragraph*{S2 Table}
\label{S2_Table}
\textbf{Parameter information of cycling model.} A list of all parameters estimated in the cycling model. Lower and upper bound denote the bound in the estimation process. Best-fit denotes the best found parameter after optimisation for one and three labels. Bounds for fatty acid influx and degradation were widened relative to the synthetic benchmark to allow these fluxes to absorb a broader range of system behaviour, since glycerol amounts were no longer explicitly modelled but instead determined by a fixed parameter.

\paragraph*{S3 Figure}
\label{S3_Figure}
\textbf{Profile likelihood analysis for all parameters in the experimental triglyceride cycling model.} Diagonal panels show profile likelihoods for selected parameters in the $L=1$ and $L=3$ models. Off-diagonal panels show re-optimised values of the row parameter as a function of the profiled column parameter, revealing compensatory parameter relationships and transitions between local optima.

\section*{Acknowledgments}

This work was supported by the Deutsche Forschungsgemeinschaft (DFG, German Research Foundation) under Germany’s Excellence Strategy (project IDs 390685813 - EXC 2047 and 390873048 - EXC 2151) and through Metaflammation, project ID 432325352 – SFB 1454, the European Union via ERC grant INTEGRATE (grant no 101126146) and by the University of Bonn via the Schlegel professorship to J.H.

\nolinenumbers

%
%

\bibliography{reference}

\end{document}